\newif\ifwordcount
\wordcountfalse

\documentclass[%
 reprint,
nofootinbib,superscriptaddress,titlepage,
 amsmath,amssymb,
prd,
]{revtex4-1}

\usepackage{color}

\usepackage{graphicx}% Include figure files
\usepackage{dcolumn}% Align table columns on decimal point
\usepackage{bm}% bold math
\usepackage{hyperref}
\usepackage{graphicx}
\usepackage{caption}
\usepackage{subcaption}
\usepackage{ragged2e}
\usepackage{multirow}
\usepackage[table,xcdraw]{xcolor}
\usepackage{hhline}
\DeclareCaptionJustification{justified}{\justifying}
\begin{document}

\title{Mitigating the optical depth degeneracy using the kinematic Sunyaev-Zel'dovich effect with CMB-S4}

\author{Marcelo A. Alvarez}
\email{marcelo.alvarez@berkeley.edu}
\affiliation{Berkeley Center for Cosmological Physics, Department of Physics,
University of California, Berkeley, CA 94720, USA}
\affiliation{Lawrence Berkeley National Laboratory, One Cyclotron Road, Berkeley, CA 94720, USA}
\author{Simone Ferraro}
\email{sferraro@lbl.gov}
\affiliation{Lawrence Berkeley National Laboratory, One Cyclotron Road, Berkeley, CA 94720, USA}
\affiliation{Berkeley Center for Cosmological Physics, Department of Physics,
University of California, Berkeley, CA 94720, USA}
\author{J.~Colin Hill}
\email{jch2200@columbia.edu}
\affiliation{Department of Physics, Columbia University, New York, NY, USA 10027}
\affiliation{Center for Computational Astrophysics, Flatiron Institute, New York, NY, USA 10010}
\author{Ren\'ee Hlo\v{z}ek}
\email{hlozek@dunlap.utoronto.ca}
\affiliation{Dunlap Institute for Astronomy and Astrophysics, University of Toronto, 50 St George Street, Toronto ON, M5S 3H4, Canada}
\affiliation{David A. Dunlap Department of Astronomy and Astrophysics, University of Toronto, 50 St George Street, Toronto ON, M5S 3H4, Canada}
\author{Margaret Ikape}
\email{ikape@astro.utoronto.ca}
\affiliation{Dunlap Institute for Astronomy and Astrophysics, University of Toronto, 50 St George Street, Toronto ON, M5S 3H4, Canada}
\affiliation{David A. Dunlap Department of Astronomy and Astrophysics, University of Toronto, 50 St George Street, Toronto ON, M5S 3H4, Canada}

\begin{abstract}
The epoch of reionization is one of the major phase transitions in the history of the universe, and is a focus of ongoing and upcoming cosmic microwave background (CMB) experiments with improved sensitivity to small-scale fluctuations.  Reionization also represents a significant contaminant to CMB-derived cosmological parameter constraints, due to the degeneracy between the Thomson-scattering optical depth, $\tau$, and the amplitude of scalar perturbations, $A_s$.  This degeneracy subsequently hinders the ability of large-scale structure data to constrain the sum of the neutrino masses, a major target for cosmology in the 2020s.  In this work, we explore the kinematic Sunyaev-Zel'dovich (kSZ) effect as a probe of  reionization, and show that it can be used to mitigate the optical depth degeneracy with high-sensitivity, high-resolution data from the upcoming CMB-S4 experiment.
We discuss the dependence of the kSZ power spectrum on physical reionization model parameters, as well as on empirical reionization parameters, namely $\tau$ and the duration of reionization, $\Delta z$. We show that by combining the kSZ two-point function and the reconstructed kSZ four-point function, degeneracies between $\tau$ and $\Delta z$ can be strongly broken, yielding tight constraints on both parameters.  We forecast $\sigma(\tau) = 0.003$ and $\sigma(\Delta z) = 0.25$ for a combination of CMB-S4 and \emph{Planck} data, including detailed treatment of foregrounds and atmospheric noise.  The constraint on $\tau$ is nearly identical to the cosmic-variance limit that can be achieved from large-angle CMB polarization data.  The kSZ effect thus promises to yield not only detailed information about the reionization epoch, but also to enable high-precision cosmological constraints on the neutrino mass.
\end{abstract}

\maketitle

\section{Introduction}
\label{sec:introduction}

The epoch of reionization (EoR) is a source of both signals and foregrounds in cosmic microwave background (CMB) observations.  The EoR is the period in cosmic history in which the baryonic contents of the Universe transitioned from a neutral to an ionized state, as a result of the ionizing radiation emitted by the first galaxies and quasars.  Along with the preceding dark ages and cosmic dawn, it is one of the least well-measured epochs in observational cosmology.  Fortunately, this situation is set to change with the advent of powerful new facilities that observe the EoR in myriad different ways, including CMB experiments (e.g., Simons Observatory~\cite{Ade:2019}, CMB-S4~\cite{CMBS4DSR}, LiteBIRD~\cite{2012SPIE.8442E..19H}), 21 cm interferometers (e.g., Hydrogen Epoch of Reionization Array~\cite{2017PASP..129d5001D}, Square Kilometer Array~\cite{2015aska.confE...1K}) and monopole experiments (e.g., EDGES~\cite{2019ApJ...875...67M}, SARAS~\cite{2018ExA....45..269S}, LEDA~\cite{2018MNRAS.478.4193P}), high-redshift galaxy surveys (e.g., Hyper Suprime-Cam~\cite{2018PASJ...70S...4A}, James Webb Space Telescope~\cite{2006NewAR..50..113W}, Roman Space Telescope~\cite{2013arXiv1305.5425S}), and many others.

In CMB measurements to date, the most relevant EoR signature has been the Thomson-scattering optical depth, $\tau = \int_{t_*}^{t_0} \bar{n}_e(t) \, \sigma_T \, dt$, where the integral runs from the surface of last scattering ($t_*$) to today ($t_0$), $\bar{n}_e$ is the cosmic-average free electron number density, and $\sigma_T$ is the Thomson cross-section.  In fact, $\tau$ is one of the six free parameters of the standard model of cosmology, $\Lambda$CDM, although unlike the others it is not a ``fundamental'' parameter of the Universe.  The optical depth predominantly influences the CMB angular power spectra in two ways: (i) the overall amplitude of the temperature and polarization power spectra on small scales is proportional to $A_s e^{-2\tau}$, where $A_s$ is the primordial amplitude of scalar fluctuations; and (ii) the large-scale ($\ell \lesssim 30$) $E$-mode polarization auto-power spectrum is proportional to $\tau^2$.  These effects arise due to the scattering of CMB photons off free electrons during the EoR, which scatters photons out of the line-of-sight (suppressing the temperature and polarization anisotropies), and generates new polarization anisotropies due to the scattering of the temperature quadrupole (analogous to the generation of $E$-mode polarization at the surface of last scattering)~(e.g.,~\cite{Zaldarriaga1997,Hu2000,Hu-Holder2003,Dore2007,Dvorkin-Smith2009}).  The latter effect is a unique signal of the EoR in the CMB power spectra, while the former effect is essentially a foreground, due to the degeneracy introduced between $A_s$ and $\tau$, which weakens constraints on the primordial amplitude.  Importantly, this also weakens constraints on beyond-$\Lambda$CDM parameters for which the sensitivity is dominated by their effect on the growth of structure between recombination and the present day, such as the sum of the neutrino masses ($M_{\nu} \equiv \sum m_{\nu}$).  Effectively, the increased error bar on $A_s$ due to the $\tau$ degeneracy becomes the limiting factor preventing a detection of $M_{\nu}$ through massive neutrinos' suppression of the growth of structure~(e.g.,~\cite{Allison2015,Pan-Knox2015,Liu2016}).  Similar degeneracies are present for dark energy and modified gravity parameters.

This situation strongly motivates measurements of $\tau$ at higher precision.  The standard method of inferring $\tau$ is via the large-angle $E$-mode power spectrum.  The current constraint from the \emph{Planck} 2018 analysis is $\tau = 0.054 \pm 0.007$~\cite{Planck2018params}, although some re-analyses have claimed error bars $\approx 30$\% smaller than this~\cite{Pagano2020}.  The ultimate cosmic variance (CV) limit on $\tau$ from the primary CMB power spectra is $\sigma(\tau) \approx 0.002$, i.e., roughly three times smaller than the \emph{Planck} error bar.  Because this signal requires measurements on the largest angular scales, it is a primary target for a next-generation satellite mission (e.g.,~LiteBIRD~\cite{2012SPIE.8442E..19H} or PICO~\cite{PICOreport}), although the Cosmology Large Angular Scale Surveyor (CLASS) is aiming to get close to this precision from the ground~\cite{2014SPIE.9153E..1IE,Watts2018}.  This gain would be significant.  If one considers CMB lensing measured by the Simons Observatory (SO) as a late-time structure growth probe, then the current \emph{Planck} $\tau$ constraint limits the neutrino mass precision to $\sigma(M_{\nu}) \approx 0.03$-$0.04$ eV, i.e., a $\lesssim 2\sigma$ detection of the minimal mass allowed by oscillation data in the normal hierarchy (0.059 eV)~\cite{Ade:2019}. If the CV limit on $\tau$ is achieved, then the identical SO CMB lensing (and CMB high-$\ell$ primary anisotropy) data would yield $\sigma(M_{\nu}) \approx 0.02$ eV, i.e., a $3\sigma$ detection of the minimal mass.  Even more significant improvements would be seen with data from CMB-S4~\cite{CMBS4DSR}.

Unfortunately, proposed satellite experiments that would reach the CV limit on $\tau$ are at least several years away from launch.  Thus, it is worth considering alternative methods with which to constrain the optical depth, which is the primary motivation for this paper.  In~\cite{Liu2016}, it was suggested that 21 cm reionization measurements could be used to constrain $\tau$.  The idea is that the 21 cm power spectrum, which traces the spatial distribution of neutral hydrogen as a function of redshift, can be used to constrain a physical model of reionization.  This model can then be used to predict $\tau$.  If the model constraints are sufficiently precise, then $\tau$ can in principle be predicted sufficiently well so as to improve on the current \emph{Planck} constraints, and eventually surpass even the CV limit from the primary CMB.

We adopt a similar approach here, but instead of the 21 cm line, we consider the kinematic Sunyaev-Zel'dovich (kSZ) effect as a probe of reionization.  The kSZ effect is the Doppler boosting of CMB photons as they Compton-scatter off free electrons moving with a non-zero velocity along the line-of-sight~\cite{1970Ap&SS...7....3S,1972CoASP...4..173S,1980ARA&A..18..537S,1986ApJ...306L..51O}.  The signal receives contributions from both the EoR, often called ``patchy'' kSZ~(e.g.,~\cite{Hu2000,Santos2003,Dore2007,Dvorkin-Smith2009}), and from galaxies, groups, and clusters at late times (sometimes called the ``homogeneous'' kSZ because the ionization fraction is essentially uniform after reionization).  The EoR kSZ signal depends sensitively on the astrophysical details of reionization, as it directly probes the distribution of free electrons.  It is effectively the complement of the 21 cm field, which directly probes the distribution of neutral hydrogen.

In this work, our primary focus is not on extracting astrophysical information from reionization kSZ measurements --- although this is a very worthwhile pursuit --- but rather on using these measurements to constrain $\tau$ and thereby resolve the parameter degeneracy problem discussed above.  We consider two statistical probes of the EoR kSZ signal: (i) the angular power spectrum (two-point function) and (ii) a particular configuration of the trispectrum (four-point function), first pointed out in~\cite{Smith:2016lnt}, with forecasts for $\tau$ presented in~\cite{Ferraro:2018izc}.
In Sec.~\ref{sec:reionkSZ}, we describe the reionization model used in this work and present the relevant two-point and four-point signals.  In Sec.~\ref{sec:experiment}, we present the CMB experiment set-up and sky modeling used in this work, including a detailed treatment of foregrounds and component separation.  Sec.~\ref{sec:results} presents our primary science results, including constraints on $\tau$ and the duration of reionization from the combination of these kSZ statistics.  We discuss these results and future challenges for this program in Sec.~\ref{sec:challenges}.

\begin{figure*}
	\includegraphics[width=0.48\textwidth]{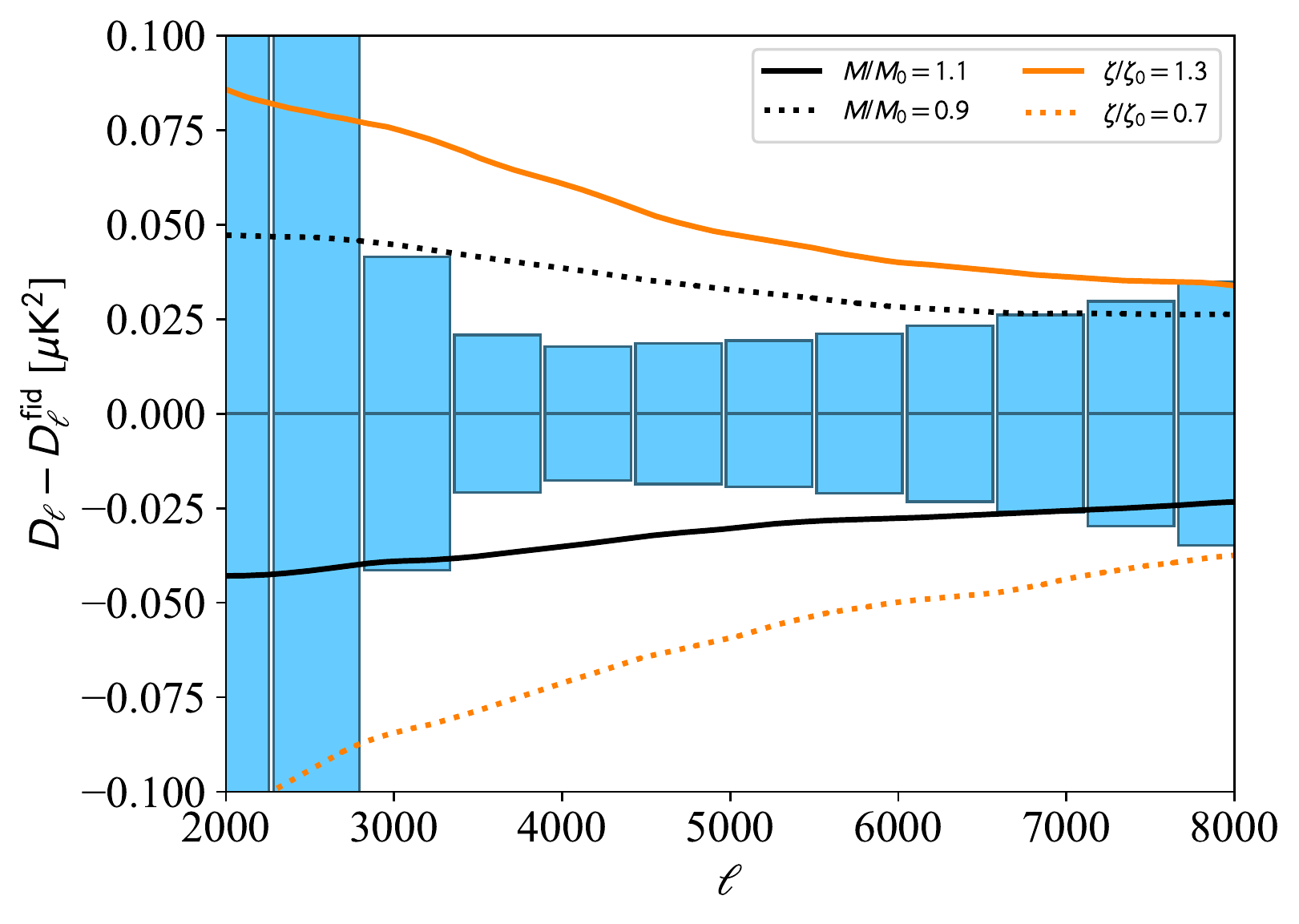}
	\includegraphics[width=0.48\textwidth]{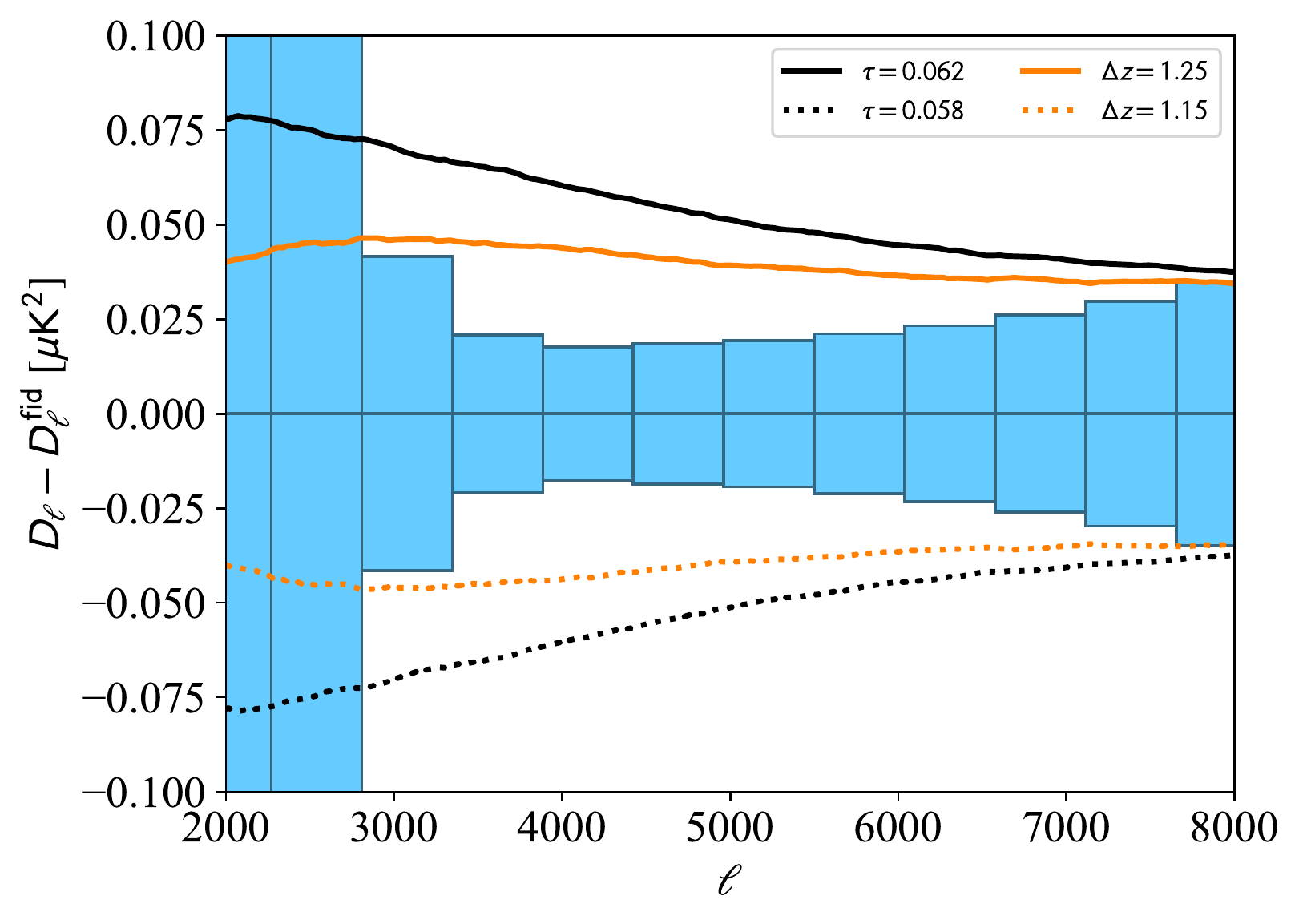}
	\caption{Dependence of kSZ power spectrum on reionization model parameters. The shaded bars show $1\sigma$ uncertainties on the power spectrum, including instrumental noise and residual foregrounds for a combination of CMB-S4 and \emph{Planck} data (see Sec.~\ref{sec:experiment}), and sample variance in the primary CMB and kSZ temperature for our fiducial model.  The solid and dotted lines show variation of input model parameters, $M_{\rm min}$ and $\zeta$ (left panel), and the resulting reionization history parameters, $\tau$ and  $\Delta z$ (right panel).  \label{fig:twopoint} 
	}
\end{figure*}

\section{Reionization kSZ}
\label{sec:reionkSZ}
%\textbf{[Marcelo]}
While the kSZ effect has long been recognized as one of the most promising probes of the intergalactic medium during and after reionization (e.g., \cite{Sunyaev1978,Kaiser1984,OV1986}), it has begun to be used only recently to provide constraints on reionization through the analysis of the angular power spectrum of the CMB temperature at $\ell \approx 3000$ \cite{Reichardt2012,Zahn2012,George2015}. A key aspect of the kSZ effect is the generation of small-scale temperature anisotropies by coupling large-scale velocity perturbations with the patchiness of the ionized field on small scales. Since, all else being equal, the kSZ power spectrum amplitude increases the earlier reionization occurs and the longer it lasts, it is an excellent probe of the reionization history.

The detailed shape and amplitude of the kSZ power spectrum varies in response to uncertain physical conditions during reionization in a complex way that can best be modeled  with simulations. 
We use the methods described in Refs.~\cite{AlvarezAbel2012} and~\cite{2016ApJ...824..118A} to simulate the reionization kSZ signal. Our simulations have as input three fundamental physical parameters controlling the morphology and history of reionization: the ionization efficiency (or number of atoms ionized per atom in halos above the minimum mass), $\zeta$; the minimum mass of halos hosting ionizing sources, $M_{\rm min}$; and the mean free path of ionizing photons, $\lambda_{\rm mfp}$.  The absorption systems that determine the mean free path are the dominant sinks of ionizing photons, limiting the size of HII regions in the percolation phase. 

The derivative of the power spectrum with respect to model parameters used in our Fisher forecasts are obtained by running a series of simulations with different parameters. Each simulation generates a realization of the ionization and density field on the observer's past light cone, from which we generate a map of the temperature fluctuation field, $(\Delta{T}/T)_{\rm kSZ}$, over 1600 square degrees, corresponding to the optical-depth-weighted line-of-sight velocity for $z>5.5$. The reionization history for each simulation on the grid of physical parameters is used to determine the Thomson scattering optical depth, $\tau$, and the duration of reionization, $\Delta{z}\equiv z_{75}-z_{25}$, the redshift interval over which the volume filling factor of ionized regions evolves from 25 to 75 percent, for each of these parameters. We also compute the power spectrum for each of these maps which, together with the mapping from physical parameters to $\tau$ and $\Delta{z}$, is used for both the two-point and four-point Fisher forecasts, as described in subsequent sections. The fiducial model values we adopt are $M_{\mathrm{min},0}=3\times 10^{9} M_{\odot}, \zeta_0 = 70$, and $\lambda_{\rm mfp, 0} = 300\ {\rm Mpc}/h$, for which $\tau_0 \simeq 0.06$ and  $\Delta z_0 \simeq 1.2$. We adopt these as fiducial values of $\tau$ and $\Delta z$ in our Fisher forecast. Fig.~\ref{fig:twopoint} illustrates the effect of varying the optical depth and duration of reionization on the kSZ power spectrum, in terms of $D_{\ell} \equiv \ell(\ell+1) C_{\ell}/(2\pi)$.

\subsection{kSZ from the Two-Point function}

As described above, we map the physical parameters controlling reionization to the empirical parameters $\tau$ and $\Delta{z}$.  Compared to template-based approaches, this model generates a kSZ power spectrum with an $\ell$ dependence. The sensitivity of upcoming CMB experiments to the temperature power on small scales, the improved ability to remove foregrounds from the power spectrum based on multi-frequency maps, and the ability of CMB polarization data to independently constrain the primary cosmological parameters all allow one to exploit this $\ell$ dependence fully. In particular, Fig.~\ref{fig:twopoint} shows the residual errors from foreground cleaning and instrumental noise (see Sec.~\ref{sec:experiment} for details) as shaded bars with simulated spectra varying the physical reionization parameters as lines. The spectral shape between $2000< \ell < 8000$ is accessible given improved sensitivity over a range of scales.

\begin{figure}
	\includegraphics[width=0.48\textwidth]{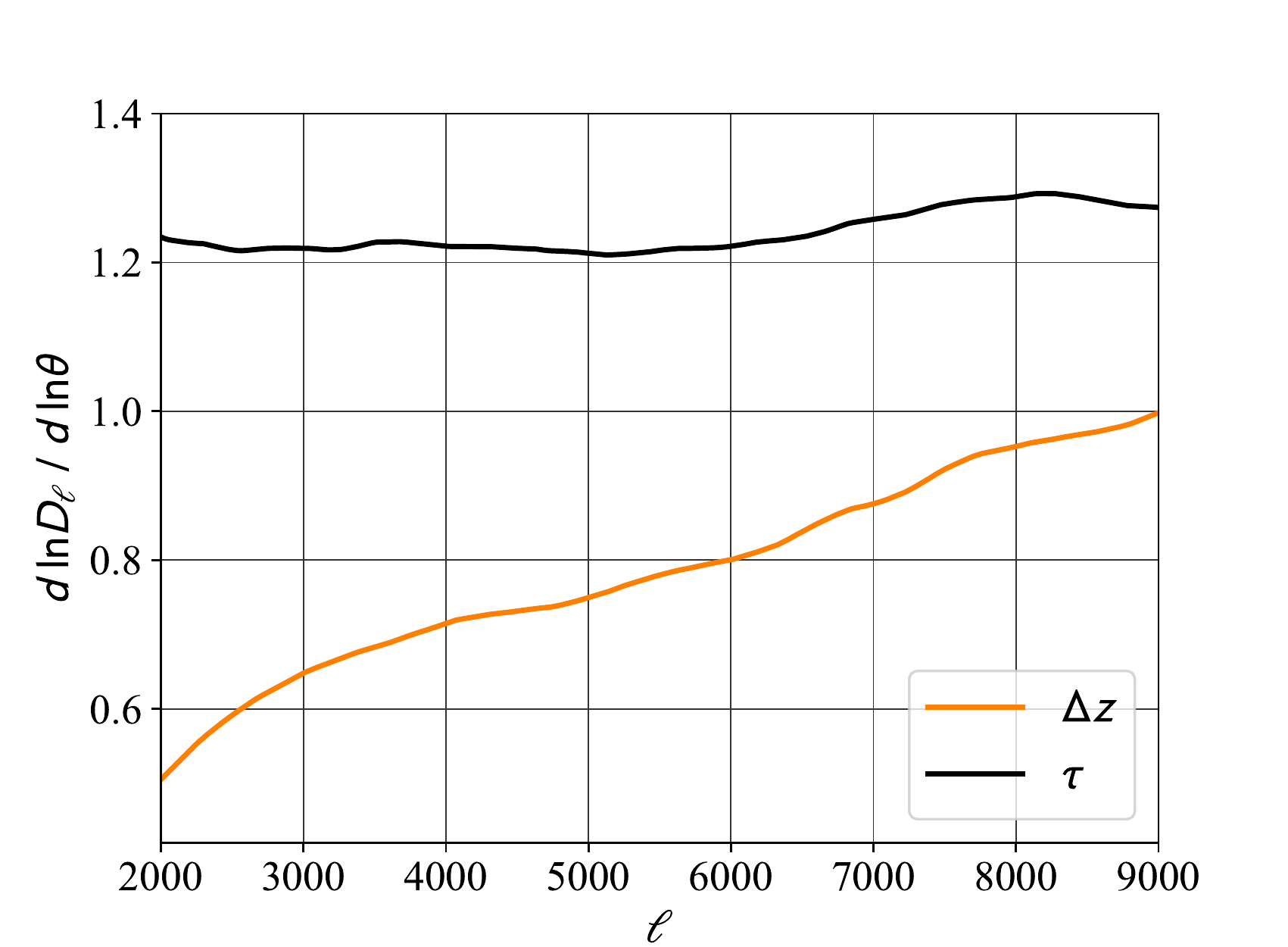}
	\caption{Response of kSZ power spectrum to variations in $\tau$ and $\Delta{z}$. Increase in either $\tau$ or $\Delta{z}$ results in more power at $\ell>1000$. The differing scale-dependence of the power spectrum response between the two parameters partially breaks the degeneracy.
	\label{fig:twopoint-derivatives}}
\end{figure}

The parameters $M_\mathrm{min}$ and $\zeta$ are the most closely related to the empirical parameters considered here, namely the optical depth $\tau$ and the duration of reionization $\Delta{z},$ and we fix $\lambda_{\rm mfp} = 300~\mathrm{Mpc}/h$ for this analysis. The dependence of the power spectrum on the parameters is illustrated in Fig.~\ref{fig:twopoint}. We compute the spectrum derivatives for the reionization parameters by varying the model parameters $M_{\rm min}$ and $\zeta$ and then use the chain rule to compute \begin{equation}
\frac{\partial C_\ell}{ \partial \tau} = \frac{\partial \zeta}{\partial \tau}\frac{ \partial C_\ell}{ \partial d\zeta} + \frac{\partial M_{\rm min}}{\partial \tau  }\frac{\partial C_\ell}{\partial M_{\rm min}} \,,
\end{equation}
and similarly for $\partial C_\ell/\partial \Delta z$.  These derivatives are shown in Fig.~\ref{fig:twopoint-derivatives}.

In the Fisher analysis, we adopt a conservative model for our prior knowledge about the late-time ``homogeneous'' kSZ contribution. We use a template for the homogeneous component from \cite{battaglia/etal:2013}, normalized to $D_\ell({\ell=3000}) = 2.0\, \mu \mathrm{K}^2.$ 
The homogeneous term can be estimated from simulations, but it is subject to astrophysical and cosmological uncertainties \cite{Shaw2012,Park2018}. 
Given the degeneracy between the homogeneous and patchy components, we do not impose strong priors on the homogeneous component. We modify the \cite{battaglia/etal:2013} template as a power law with a pivot at $\ell=3000,$ with an amplitude and slope with fiducial values of $A_\mathrm{homKSZ}=1,\alpha_\mathrm{homKSZ}=0 $. We marginalize over both terms with a flat, non-informative prior. The constraints are not strongly dependent on the choice of prior for the homogeneous parameters; the homogeneous parameters are constrained by the data to $\sigma(A_\mathrm{homKSZ}) = 0.42, \sigma(\alpha_\mathrm{KSZ}) = 0.48$.  However, imposing a 10\% prior on the homogeneous amplitude improves the error on the duration of reionization by 25\%, as discussed in Section~\ref{sec:results}. We treat the optical depth inferred from the reionization kSZ signal and the optical depth inferred from the primary CMB as separate parameters, allowing both to vary and marginalizing over the primary $\tau_\mathrm{CMB}$, as our goal here is to isolate the reionization information coming from the kSZ signal alone. For the two-point forecasts we include the temperature, polarization, and cross-power spectra (TT, EE, TE). To be conservative as to any residual foregrounds that persist after multi-frequency cleaning, we restrict the TT power spectrum to $30<\ell<3000$ and use the TE and EE power spectrum between $30 < \ell < 5000,$ following a similar treatment to that presented in \cite{Ade:2019}. Details regarding the foreground and noise models are given in Sec.~\ref{sec:experiment}. Pushing to higher $\ell$ in TT could significantly improve the constraints derived from the two-point function, but would require a very accurate model for the CIB, tSZ, and other small-scale foregrounds in order to avoid biases.

\subsection{kSZ from the Four-Point function}
In addition to being the largest blackbody component in the high-$\ell$ CMB, the kSZ signal is also significantly non-Gaussian.  This is because small-scale fluctuations in the ionization fraction are modulated by slowly varying velocity fields, meaning that the locally-measured kSZ power spectrum varies significantly between different sightlines with different realizations of the velocity. Such a modulation can be detected by a four-point function estimator \cite{Smith:2016lnt, Ferraro:2018izc}, in close analogy to the one used to reconstruct the CMB lensing power spectrum.
Another feature of this estimator is that the \textit{shape} of the measured four-point function is determined by the properties of the velocity field, which is well described by linear theory. The velocity coherence length acts as a ``standard ruler'', allowing us to separate the late-time and reionization contributions to the kSZ signal in a model-independent way \cite{Smith:2016lnt, Ferraro:2018izc}.

In addition to the derivatives of $C_\ell$ with respect to the parameters $(\tau, \Delta z)$, for the four-point function analysis, we also need to assume a redshift distribution of the source of the reionization kSZ signal.  Following \cite{ Ferraro:2018izc}, we take
\begin{equation}
    \left( \frac{d C_{\ell}}{d z} \right)_{\rm rei}(z, l, \tau, \Delta z) = C_{\ell, \mathrm{rei}}(\tau, \Delta z) \frac{e^{-(z-\bar{z})^2 / 2 \sigma_z^2}}{\sqrt{2 \pi \sigma_z^2}}
\label{eq:rei_der}
\end{equation}
where we take the duration $\Delta z$ to be approximately the full-width at half maximum (FWHM) of the distribution, such that $\sigma_z \approx \Delta z / \sqrt{8 \ln 2}$, and $\bar{z}(\tau)$ is the mean redshift of reionization.

In this paper, we use the forecasting formalism of \cite{Ferraro:2018izc}, marginalizing over an arbitrary amplitude and shape for the late-time kSZ (with no prior) as well as a white noise contribution.  
For the purpose of this paper, we define the reionization contribution as being all of the kSZ signal coming from $z > 6$. Since the bulk of the late-time kSZ originating from galaxies and clusters originates from much lower redshift, we find that our results are insensitive to this particular choice. In addition to providing robustness in separating the late-time component, the kSZ four-point function has a different parameter dependence than the power spectrum, allowing for very effective degeneracy breaking, which is the main result of this work (see Sec.~\ref{sec:results}).

Gravitational lensing of the CMB creates a four-point function that could potentially mimic that of reionization.  In \cite{Smith:2016lnt} it was shown that using lensing reconstruction from polarization (which is not affected by kSZ), the lensing contribution can be reduced to a white noise component, which we marginalize over in the forecast here.  We use temperature modes from $2000 < \ell < 6000$ in the four-point forecast.

\section{Experimental assumptions and sky modeling}
\label{sec:experiment}

In this work, we consider forecasts for the future CMB-S4 experiment~\cite{CMBS4DSR} (first light $\sim 2027$), which has sufficient sensitivity and resolution to enable robust application of our methods.  While ongoing ground-based CMB surveys (e.g., Advanced ACT~\cite{Henderson:2016} and SPT-3G~\cite{Benson:2014}) will also measure the kSZ power spectrum (including reionization contributions), their sensitivity is not sufficient to measure the kSZ four-point function~\cite{Smith:2016lnt, Ferraro:2018izc}, which is crucial for breaking parameter degeneracies, as discussed in detail below.  The SO nominal survey~\cite{Ade:2019} (first light $\sim 2022$) may have sufficient sensitivity to make a first detection of the kSZ four-point function, but the signal-to-noise on this statistic is likely to be too low to enable the parameter degeneracy-breaking described below.

The experimental specifications of the CMB-S4 survey used here are presented in detail in Ref.~\cite{CMBS4DSR}.  We focus solely on the high-resolution, wide-area CMB-S4 survey (the experiment will also include a low-resolution, small-area, ultra-deep survey for primordial B-modes).  Summarizing the setup briefly, the two CMB-S4 large aperture telescopes (LATs) used for the wide-area survey will include six frequency channels centered at 27, 39, 93, 145, 225, and 280 GHz.  The CMB-S4 LATs will employ diffraction-limited optics on telescopes with a 6-meter primary dish, yielding a beam with FWHM = 1.4 arcmin at 145 GHz (and scaling inversely with frequency).  Complete details of the CMB-S4 noise modeling --- including both instrumental noise and non-white atmospheric noise with realistic frequency dependence --- are located in Ref.~\cite{CMBS4DSR}.  As an approximate guide, the anticipated high-multipole white noise level of the CMB-S4 wide-field survey is 2~$\mu$K$\cdot$arcmin at 93 GHz and 2~$\mu$K$\cdot$arcmin at 145 GHz.  This survey will encompass 70\% of the sky, but we assume an effective sky area of 45\% for the high-precision CMB blackbody temperature map reconstruction that is a necessary first step for the kSZ analyses considered below (our forecasts are thus somewhat conservative).

To forecast an effective post-component-separation noise power spectrum for the reconstructed CMB blackbody temperature map, we employ the methodology described in Ref.~\cite{CMBS4DSR} (see Appendix A.3; see also Sec.~2 of Ref.~\cite{Ade:2019}).  \emph{Planck} data from 30--353 GHz are also assumed to be used in the CMB blackbody component separation; these data are particularly crucial on large angular scales where atmospheric noise is significant for CMB-S4 and other ground-based experiments.  Nevertheless, we emphasize that the forecasts here are driven by the high-sensitivity, multi-frequency data of CMB-S4 on small angular scales, where the kSZ signal dominates the blackbody sky.  In total, we consider thirteen frequency channels, six from CMB-S4 and seven from \emph{Planck}.  Our component separation analysis includes realistic models of all major sky signals and foregrounds for every \emph{Planck} and CMB-S4 frequency channel, combined with the CMB-S4 noise modeling mentioned above and white noise for the \emph{Planck} channels (with noise levels from~\cite{Planck2015LFImapmaking,Planck2015HFImapmaking}).  We then analyze these sky models with a harmonic-space internal linear combination (ILC)~\citep[e.g.,][]{Eriksen2004} code to compute post-component-separation noise power spectra for the cleaned CMB blackbody temperature map, $N_{\ell}^{TT}$.  These power spectra thus capture the contributions of residual foregrounds and noise due to the detectors and atmosphere.  %We consider the multipole range from $80 < \ell < 8000$ in this work.

For simplicity, we use ``standard'' ILC noise power spectra here, in which the total variance of the final blackbody map is minimized (subject to a constraint that preserves the signal), but in which no particular contaminant is explicitly required to vanish.  Future analyses may necessitate the use of CMB blackbody ILC maps with particular component SEDs nulled (e.g., tSZ or approximate CIB SEDs) via a constrained ILC procedure~\citep[e.g.,][]{Remazeilles2011,MHN2019} so as to mitigate possible biases from these contaminants.  It may also be the case that the kSZ power spectrum will be inferred through an analysis directly at the power spectrum level, i.e., without first constructing a foreground-cleaned blackbody map.  (Measuring the kSZ four-point function will almost certainly require constructing a foreground-cleaned map first.)   Both of these analysis choices could modestly increase the error bars on the forecasts presented here.  High-frequency maps from, e.g., CCAT-prime~\cite{CCAT} could be useful in mitigating foreground contamination effects, particularly due to the cosmic infrared background (CIB).  Due to current uncertainties in CIB modeling, we defer detailed consideration of this issue to future work employing an end-to-end map-based simulation framework.

\begin{figure}
	\includegraphics[width=0.5\textwidth]{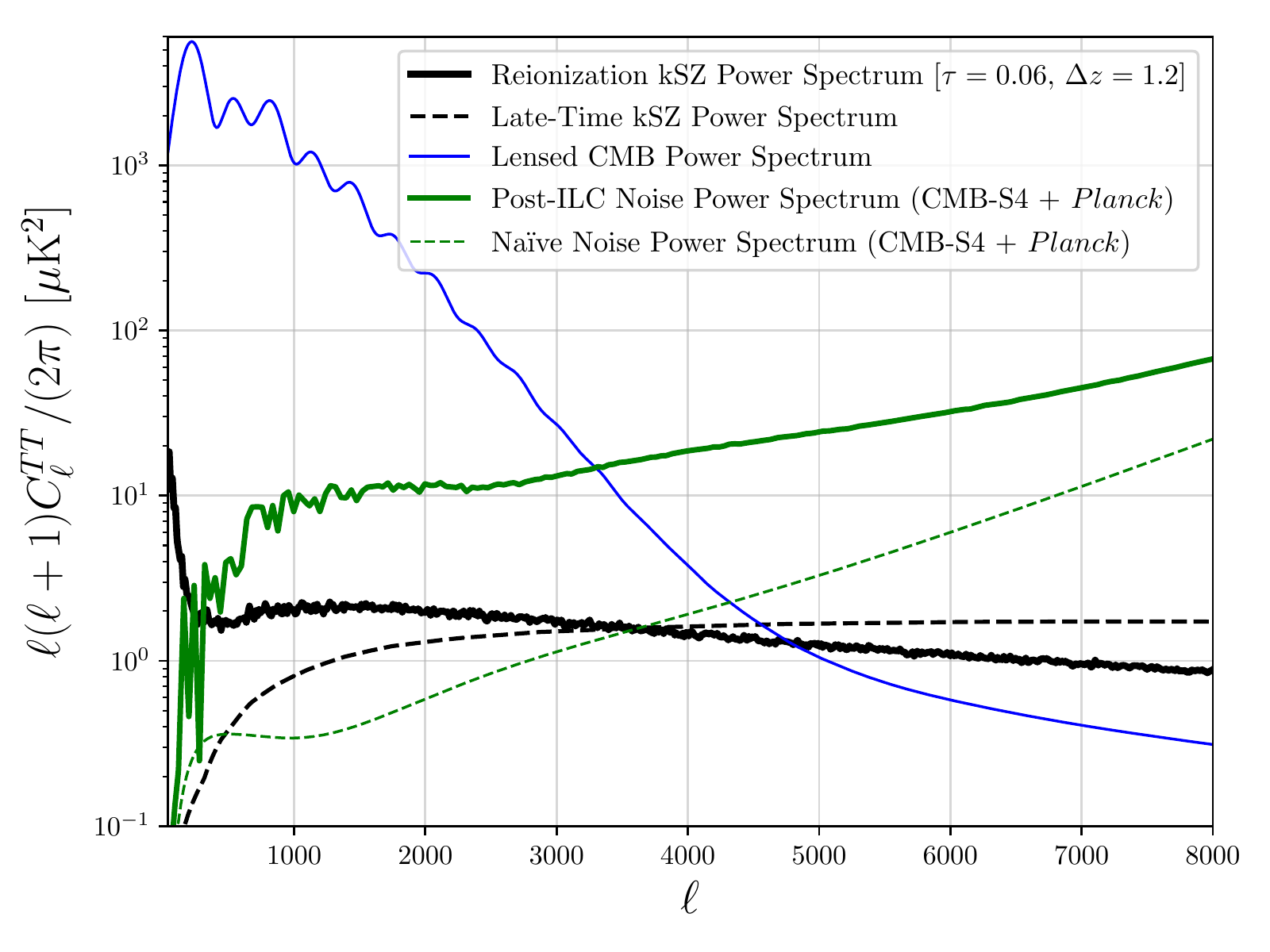}
	\caption{Signal and noise power spectra.  The thick black curve shows our fiducial reionization kSZ power spectrum, computed in a model with $\lambda_{\rm mfp} = 300$ Mpc/$h$, $\zeta = 70$, and $M_{\rm min} =3 \times 10^9 \, M_{\odot}$, which yields $\tau = 0.06$ and $\Delta z = 1.2$.  The dashed black curve shows our fiducial late-time kSZ power spectrum, while the thin blue curve shows the lensed primary CMB temperature power spectrum.  Both of these contributions are spectrally degenerate with the reionization kSZ signal.  The thick green curve shows the effective noise power spectrum determined from CMB-S4 and \emph{Planck} data using an ILC method, while the dashed green curve shows the na\"{i}ve noise power spectrum in the absence of foregrounds.}
	\label{fig:noise}
\end{figure}

Fig.~\ref{fig:noise} shows the final post-component-separation noise power spectrum used in this analysis, as well as the CMB blackbody signal comprised of the lensed primary temperature power spectrum and the kSZ power spectrum.  The latter includes contributions from both reionization and the late-time universe, as labeled in the figure.  For comparison, the figure also shows a na\"{i}ve noise power spectrum that would result if all of the frequency maps were co-added with inverse-noise-variance weighting only, and no foregrounds were present in the sky.  This highlights the importance of fully modeling all signals in the mm-wave sky in such forecasts.

\begin{figure*}[htbp!]
    \centering
\includegraphics[width=0.7\textwidth]{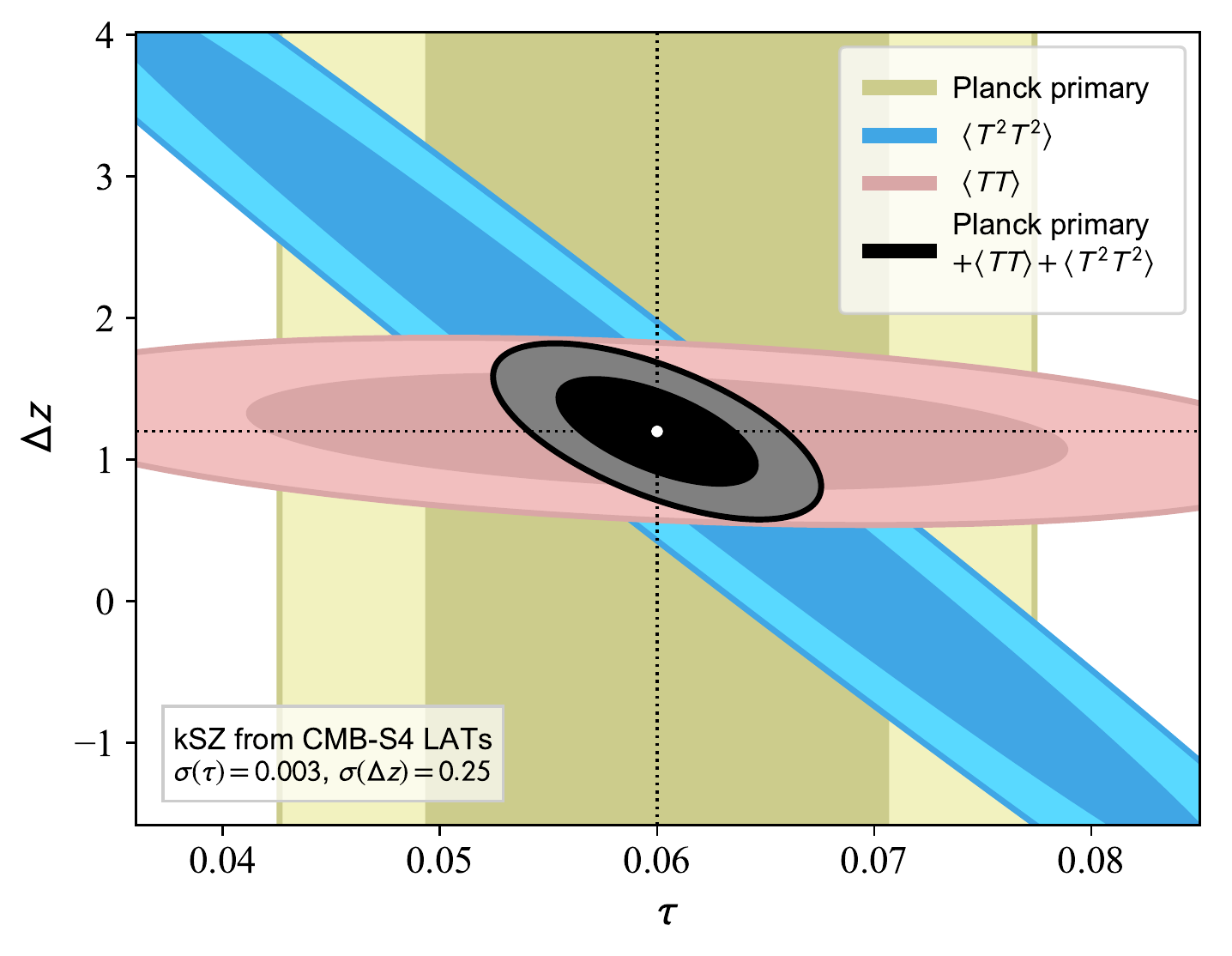}
	\caption{Constraints on the duration of reionization and optical depth. The vertical shaded contours are 68\% and 95\% confidence regions from the primary CMB anisotropies measured by \emph{Planck}, which constrain the optical depth to an error of $\sigma(\tau)=0.007$ (the primary CMB does not constrain $\Delta{z}$).  The angled contours show forecast reionization constraints from the kSZ power spectrum (pink) and the kSZ four-point function (blue), as derived from CMB-S4 and \emph{Planck} data.  The black contours show forecast constraints from the combination of all three probes.  The complementary degeneracy directions of the two-point and four-point functions effectively break the degeneracy between the reionization parameters, yielding tight constraints on both $\tau$ and $\Delta z$: $\sigma(\tau) = 0.003$ and $\sigma(\Delta z) = 0.25$.} 
	\label{fig:constraints}
\end{figure*}

\section{Results}
\label{sec:results}
%\textbf{[Everybody]}

We show the individual and joint constraints on $\Delta z$ and $\tau$ from the kSZ two-point and four-point functions in Fig.~\ref{fig:constraints}. The two-point function is weakly constraining on the optical depth compared to the standard constraints from the primary CMB, but tightly constrains the duration of reionization. Conversely, the four-point function is more sensitive to the optical depth than to the duration of reionization. When combined together, and also folding in the \emph{Planck} primary CMB constraint on $\tau$, the joint forecast yields a (marginalized) covariance matrix
\begin{equation*}
    {\rm Cov}(\tau, \Delta z) = 
    \begin{pmatrix}
    9.3\times{10}^{-6} & -4.7\times{10}^{-4} \\
    -4.7\times{10}^{-4} & 0.063
    \end{pmatrix}
\end{equation*}
so that  $\sigma(\tau) = 3\times{10}^{-3}$ and $\sigma(\Delta z) = 0.25$.  The error on $\sigma(\Delta z)$ reduces to $\sigma(\Delta z) = 0.2$ if we place a 10\% prior on the amplitude of the homogeneous signal. This constraint on $\tau$ is nearly as tight as a CV-limited constraint from the primary CMB ($\sigma(\tau) = 2\times{10}^{-3}$), as targeted by next-generation satellite missions. 

The different degeneracy direction between the two- and four-point estimators is straightforward to explain: while changing parameters such as the duration of reionization changes the power spectrum, it will also change the amount of non-Gaussianity in a different way. For example, a shorter reionization epoch would lead to a more non-Gaussian kSZ field, and enhance the four-point function compared to the two-point function.  By measuring both, we can effectively break the parameter degeneracy and obtain tighter limits on reionization.

Since the four-point estimator involves four powers of the map noise, one may wonder whether it would perform better in a deeper but smaller survey (e.g., the ``delensing'' survey planned for the CMB-S4 primordial gravitational wave search~\cite{CMBS4DSR}), rather than in the shallower wide survey considered here. A simple estimate indicates that because of foregrounds, the reduction in effective noise is not large enough to compensate for the decreased sky area, and thus the wide survey considered here is expected to yield better performance.

\section{Discussion and Challenges}
\label{sec:challenges}
Fig.~\ref{fig:constraints} clearly illustrates the power of combining the four-point and two-point constraints due to their complementary degeneracy directions in the reionization parameter space.  Given this statistical power, a careful consideration of potential biases and systematics of these probes is necessary, which we briefly outline here.

\textit{Foreground cleaning:} Multi-frequency coverage is crucial for isolating the blackbody kSZ signal from other, non-blackbody foregrounds in the high-$\ell$ CMB, such as the thermal SZ effect and the CIB.  At the power spectrum level, these contributions can be simultaneously fit in a multi-component analysis, although accurate modeling will be needed.  To be conservative in this work, we have not used modes at $\ell>3000$ in the two-point function forecast.  For the four-point function, it is likely optimal to first construct a foreground-cleaned blackbody map before measuring the statistic. The late-time kSZ contribution is explicitly marginalized out in the four-point analysis, but it is important to have a sufficiently flexible model of the late-time power spectrum to marginalize over in the two-point analysis. Additional constraints from cross-correlations with spectroscopic galaxy surveys will also help to reduce uncertainty associated with the late-time contribution to the power spectrum. Precisely calibrating residual biases in either estimator due to foreground leakage or mismodeling will require dedicated simulations, but this should not be an insurmountable obstacle on the timescale of CMB-S4. 

\textit{Reionization modelling:} Perhaps the largest source of uncertainty is the physical modelling of the reionization process in a standard UV-dominated scenario, and more specifically in the parameter dependence of the mean free path, efficiency, and mass. Alternative  reionization scenarios involving very high-redshift sources would involve different values of the parameters and model assumptions than those considered here. We leave the investigation of the sensitivity to these models to future work; however, the precision of the model parameter constraints in this analysis implies that we will indeed be able to rule out other models including reionization from early X-ray binaries, population~III sources, rare quasars, or other exotic reionization scenarios.  Also, our ability to pin down the exact model of reionization will be enhanced through cross-correlations of CMB-S4 data with external data sets such as 21-cm and Lyman-$\alpha$ emitter surveys~(e.g.,~\cite{2020arXiv200507206L}). Finally, we note that independent CMB-based constraints on the reionization history and optical depth $\tau$ from the large-scale $EE$ power spectrum measurements will further break the degeneracy by removing uncertainty on one axis.

\textit{Non-patchy optical depth:}
Here we have assumed that all the patchiness in the ionization field will be resolved by the measurements. However, certain scenarios, such as reionization due to dark matter decay or annihilation or very hard X-ray sources, allow for additional contributions to the optical depth that do not contribute patchiness (or the bubbles are too small to be resolved given the finite beam size). Thus, these constraints are technically a lower limit on the optical depth. 

\textit{Covariance:}
In the analysis above, we assumed that the kSZ two-point and four-point functions had zero covariance.  This assumption holds if the patches in which these signals are measured are non-overlapping on the sky, in which case the noise covariance is clearly zero and we can straightforwardly combine them. If they are overlapping, the calculation involves computing $\sim$ 400 six-point functions, and we leave it to future work (a simulation-based analysis may be more tractable). The uncorrelated assumption also holds if we use two different experiments on the same patch of sky, since our constraints are dominated by high-$\ell$ information where the primary CMB is negligible.

While overcoming the challenges mentioned above will require significant effort, this is well justified by the kSZ reionization constraints forecast here.  The tight constraint on $\tau$ will enable neutrino mass constraints from upcoming surveys that utilize the full statistical power available from large-scale structure data, including CMB lensing.  The reionization constraints will yield rich astrophysical information about the nature and distribution of the ionizing sources, in particular when kSZ data are jointly analyzed with 21 cm data, intensity mapping surveys, high-redshift galaxy and quasar studies, and other probes of the EoR.

\section*{Acknowledgments}
We thank Tom Crawford, Emmanuel Schaan, Blake Sherwin and Kendrick Smith for useful discussions.  SF is supported by the Physics Division of Lawrence Berkeley National Laboratory. JCH thanks the Simons Foundation for support. RH is a CIFAR Azrieli Global Scholar, Gravity \& the Extreme Universe Program, 2019, and a 2020 Alfred P. Sloan Research Fellow. RH is supported by Natural Sciences and Engineering Research Council of Canada. The Dunlap Institute is funded through an endowment established by the David Dunlap family and the University of Toronto. We acknowledge that the land on which the University of Toronto is built is the traditional territory of the
Haudenosaunee, and most recently, the territory of the Mississaugas of the New Credit First Nation. We are grateful
to have the opportunity to work in the community, on this
territory. All authors contributed equally to the preparation of this manuscript.

\bibliographystyle{apsrev}
\bibliography{reionkSZ}
%\fi
%\begin{thebibliography}{99}

%\end{thebibliography}
\end{document}